\documentclass[aps,twocolumn,prb,showpacs,superscriptaddress,floatfix]{revtex4}
\usepackage{amsmath} \usepackage{graphicx} \usepackage{amssymb}
\usepackage{bm}

\newcommand{\e}{\mathrm{e}}

\DeclareMathAlphabet{\mathitbf}{T1}{cmr}{bx}{it} 

\begin{document}
  
\title{Critical Behavior of Three-Dimensional Disordered Potts Models
  with Many States}

\author{R. Alvarez Ba\~nos}\affiliation{Departamento
  de F\'\i{}sica Te\'orica, Universidad
  de Zaragoza, 50009 Zaragoza, Spain.}
  \affiliation{Instituto de Biocomputaci\'on y
  F\'{\i}sica de Sistemas Complejos (BIFI), Zaragoza, Spain.}

\author{A.~Cruz} \affiliation{Departamento
  de F\'\i{}sica Te\'orica, Universidad
  de Zaragoza, 50009 Zaragoza, Spain.} 
  \affiliation{Instituto de Biocomputaci\'on y
  F\'{\i}sica de Sistemas Complejos (BIFI), Zaragoza, Spain.}

\author{L.~A.~Fernandez} \affiliation{Departamento
  de F\'\i{}sica Te\'orica I, Universidad
  Complutense, 28040 Madrid, Spain.} 
  \affiliation{Instituto de Biocomputaci\'on y
  F\'{\i}sica de Sistemas Complejos (BIFI), Zaragoza, Spain.}

\author{A.~Gordillo-Guerrero}\affiliation{Dpto. de Ingenier\'{\i}a
  El\'ectrica, Electr\'onica y Autom\'atica,  Universidad de Extremadura,
  10071 C\'aceres, Spain.}
\affiliation{Instituto de Biocomputaci\'on y F\'{\i}sica de Sistemas
  Complejos (BIFI), Zaragoza, Spain.}

\author{J.~M.~Gil-Narvion} \affiliation{Departamento
  de F\'\i{}sica Te\'orica, Universidad
  de Zaragoza, 50009 Zaragoza, Spain.}
  \affiliation{Instituto de Biocomputaci\'on y
  F\'{\i}sica de Sistemas Complejos (BIFI), Zaragoza, Spain.}

\author{M.~Guidetti} \affiliation{Dipartimento
  di Fisica, Universit\`a di Ferrara and INFN - Sezione di Ferrara,
  Ferrara, Italy.} 

\author{A.~Maiorano} \affiliation{Dipartimento di Fisica, CNR and INFN,
Universit\`a di Roma ``La Sapienza'', 00185 Roma, Italy.}

\author{F.~Mantovani} \affiliation{Dipartimento
  di Fisica, Universit\`a di Ferrara and INFN - Sezione di Ferrara,
  Ferrara, Italy.} 

\author{E.~Marinari} \affiliation{Dipartimento di Fisica, CNR and INFN,
Universit\`a di Roma ``La Sapienza'', 00185 Roma, Italy.}

\author{V.~Martin-Mayor} \affiliation{Departamento de F\'\i{}sica
  Te\'orica I, Universidad Complutense, 28040 Madrid, Spain.} 
\affiliation{Instituto de Biocomputaci\'on y
  F\'{\i}sica de Sistemas Complejos (BIFI), Zaragoza, Spain.}

\author{J.~Monforte-Garcia} \affiliation{Departamento
  de F\'\i{}sica Te\'orica, Universidad
  de Zaragoza, 50009 Zaragoza, Spain.}
  \affiliation{Instituto de Biocomputaci\'on y
  F\'{\i}sica de Sistemas Complejos (BIFI), Zaragoza, Spain.}

\author{A.~Mu\~noz Sudupe} \affiliation{Departamento
  de F\'\i{}sica Te\'orica I, Universidad
  Complutense, 28040 Madrid, Spain.} 

\author{D.~Navarro} \affiliation{Departamento  de Ingenier\'{\i}a,
  Electr\'onica y Comunicaciones and Instituto de Investigaci\'on en\\
  Ingenier\'{\i}a de Arag\'on (I3A), Universidad de Zaragoza, 50018 Zaragoza, Spain.}

\author{G.~Parisi} \affiliation{Dipartimento di Fisica, CNR and INFN,
Universit\`a di Roma ``La Sapienza'', 00185 Roma, Italy.}

\author{S.~Perez-Gaviro} 
\affiliation{Dipartimento di Fisica, CNR and INFN,
Universit\`a di Roma ``La Sapienza'', 00185 Roma, Italy.}
\affiliation{Instituto de Biocomputaci\'on y
  F\'{\i}sica de Sistemas Complejos (BIFI), Zaragoza, Spain.}

\author{J.~J.~Ruiz-Lorenzo} \affiliation{Departamento de
  F\'{\i}sica, Universidad de Extremadura, 06071 Badajoz, Spain.}
\affiliation{Instituto de Biocomputaci\'on y
  F\'{\i}sica de Sistemas Complejos (BIFI), Zaragoza, Spain.}

\author{B.~Seoane}  \affiliation{Departamento de F\'\i{}sica
  Te\'orica I, Universidad Complutense, 28040 Madrid, Spain.}
  \affiliation{Instituto de Biocomputaci\'on y
  F\'{\i}sica de Sistemas Complejos (BIFI), Zaragoza, Spain.}

\author{S.~F.~Schifano} \affiliation{Dipartimento di
  Matematica, Universit\`a di Ferrara and INFN - Sezione di Ferrara,
  Ferrara, Italy.} 

\author{A.~Tarancon} \affiliation{Departamento
  de F\'\i{}sica Te\'orica, Universidad
  de Zaragoza, 50009 Zaragoza, Spain.} 
  \affiliation{Instituto de Biocomputaci\'on y
  F\'{\i}sica de Sistemas Complejos (BIFI), Zaragoza, Spain.}

\author{R.~Tripiccione} \affiliation{Dipartimento
  di Fisica, Universit\`a di Ferrara and INFN - Sezione di Ferrara,
  Ferrara, Italy.} 
 
\author{D.~Yllanes}  \affiliation{Departamento de F\'\i{}sica
  Te\'orica I, Universidad Complutense, 28040 Madrid, Spain.}
  \affiliation{Instituto de Biocomputaci\'on y
  F\'{\i}sica de Sistemas Complejos (BIFI), Zaragoza, Spain.}

\date{\today}

\begin{abstract}
  We study the $3D$ Disordered Potts Model with $p=5$ and $p=6$.  Our
  numerical simulations (that severely slow down for increasing $p$)
  detect a very clear spin glass phase transition. We evaluate the
  critical exponents and the critical value of the temperature, and we
  use known results at lower $p$ values to discuss how they evolve for
  increasing $p$. We do not find any sign of the presence of a transition
  to a ferromagnetic regime.
\end{abstract}

\pacs{75.50.Lk, 75.40.Mg, 64.60.F-}

\maketitle

\section{Introduction}

The three dimensional ($3D$) disordered Potts model (DPM) is an
important system, that could help in clarifying a number of open and
crucial questions.  The first issue that comes to the mind is the
possibility of understanding the glass transition, since this is a
very challenging problem. On more general grounds, it is very
interesting to try and qualify the behavior of the system when the
number of states $p$ becomes large: here we should see the paradigm of
a ``hard'', first order like transition but, as we will discuss in the
following, only sometimes this turns out to be clear (see for example
the set of large scale, very accurate numerical simulations
of Ref. ~\onlinecite{BBK}, dealing with a model slightly different from the
one defined here).

In such a difficult situation extensive numerical simulations are more than
welcome, and the Janus supercomputer~\cite{JANUS2,JANUS08}, optimized for
studying spin glasses, reaches its peak performances when analyzing lattice
regular systems based on variables that can take a finite, small number of
values: disordered Potts models fit very well these requirements.  Using the
computational power of Janus we have been able to consistently thermalize the
DPM with $p=5$ and $6$ on $3D$ (simple cubic) lattice systems with periodic
boundary conditions and size up to $L=12$. Bringing these systems to thermal
equilibrium becomes increasingly harder with increasing number of states: it
has been impossible for us, even by using a large amount of time of Janus
(that for these problems performs, as we discuss better in the following, as
thousands of PC processors), to get a significant, unbiased number of samples
thermalized, and reliable measurements of physical quantities, for $p\ge 5$ on
a $L=16$ lattice.

Our results lead us to the claim that the critical behavior of the DPM
with a large number of states $p$ is very subtle, and if $p$ is larger
than, say, $5$, numerical simulations could easily give misleading
hints. The numerical results that we will discuss in the following
lead us to believe that the spin glass transition gets stronger with
increasing number of states $p$: a theoretical analysis of these
results suggests that the transition could eventually become of first
order for $p$ large enough. We do not observe, for both $p=5$ and
$p=6$, any sign of the presence of a spontaneous magnetization.

\section{Model and observables}\label{sec:MODEL}

We have performed numerical simulations of the DPM on a simple cubic
lattice of linear size $L$ with periodic boundary conditions. The
Hamiltonian of the DPM is
\begin{equation}
  \label{eq:ham}
  {\cal H} \equiv - \sum_{\langle i,j \rangle}J_{ij}\, \delta_{s_i,s_j}\,,
\end{equation}
where the sum is taken over all pairs of first neighboring sites. In
the $p$-states model spins $s_i$ can take $p$ different values
$\{0,1, \ldots,\,p-1\}$. In this work we analyze the $p=5$ and $6$
cases.  The couplings $J_{ij}$ are independent random variables taken
from a bimodal probability distribution ($J_{ij}=\pm 1$ with
probability $\frac12$). For a different definition of a disordered
Potts model see Ref.~\onlinecite{MMP}.

It is convenient to rewrite the variables of the Potts model using the
{\it simplex} representation, where the $p$ Potts states are
described as vectors pointing to the corners of a $(p-1)$
dimensional hyper-tetrahedron. The Potts scalar spins $s_i$ are thus
written as $(p-1)$-dimensional unit vectors $\mathitbf{S}_i $
satisfying the relations
\begin{equation}
  \label{eq:simplex}
  \mathitbf{S}_a \cdot \mathitbf{S}_b = \frac{p\,\delta_{a b}-1}{p-1}\, ,
\end{equation}
where $a$ and $b$ $\in [1,p]$.  We use this vector representation to
define the observables required to investigate the critical behavior
of the system.  In the simplex representation we have that:
\begin{equation}
 \label{eq:ham_simplex}
H = - \sum_{\langle i,j \rangle}J'_{ij}\, 
\mathitbf{S}_i \cdot
\mathitbf{S}_j\, .
\end{equation}
The couplings in the simplex representation
have the form
\begin{equation}
 \label{eq:JnewJold}
J'_{ij} = \frac{p-1}{p}  J_{ij} \, .
\end{equation}
The spin glass behavior is studied via a properly defined tensorial
overlap between two \emph{replicas} (independent copies of the system
characterized by the same quenched disorder variables $J_{ij}$). Its
Fourier transform (with wave vector $\mathitbf{k}$) is
given by~\cite{KLY}
\begin{equation}
  \label{eq:overlap}
  q^{\mu \nu}(\mathitbf{k}) = \frac{1}{V} \sum_i 
  S_{i}^{(1)\mu} S_{i}^{(2)\nu} \e^{i \mathitbf{k} \cdot \mathitbf{R}_i}\,,
\end{equation}
where ${S}_{i}^{(1)\mu}$ is the $\mu$ component of the spin at
site $i$ of the first replica in the simplex representation,
${S}_{i}^{(2)\nu}$ the $\nu$ component of the spin at site $i$ in
the second replica, and $V=L^3$ is the volume of the system.

This spin glass order parameter is then used to define the
spin glass susceptibility in Fourier space.
\begin{equation}
  \label{eq:chisg}
  \chi_{q}(\mathitbf{k}) \equiv V \sum_{\mu,\nu}
  \overline{\langle |q^{\mu \nu} (\mathitbf{k})|^2 \rangle}\,,
\end{equation}
where $\langle (\cdot\cdot \cdot) \rangle$ indicates a thermal average and
$\overline{(\cdot\cdot\cdot)}$ denotes the average over different realizations
of the disorder (\emph{samples} in the following). With the above definition, $\chi_q(0)$
is the usual spin glass susceptibility.

We are interested in studying the value of the dimensionless
correlation length $\xi/L$, since at the transition temperature it
does not depend on $L$, and is therefore extremely helpful to estimate
the critical temperature value $T_c$: in fact one can usually simulate
different lattice sizes, and look for the crossing point in the plot
of the different $\xi/L$ values.  One can derive~\cite{VICTORAMIT} the
value of the correlation length $\xi$ from the Fourier transforms of
the susceptibility with
\begin{equation}
  \label{eq:xi}
  \xi = \frac{1}{2\sin{(\mathitbf{k}_{\mathrm{m}}/2)}} 
  \bigg( \frac{\chi_q(0)}{\chi_q(\mathitbf{k}_{\mathrm{m}})} 
  - 1 \bigg)^{1/2}\,,
\end{equation}
where $\mathitbf{k_{\mathrm{m}}}$ is the minimum wave vector allowed
in the lattice. With the periodic boundary conditions used in this
work we have $\mathitbf{k_{\mathrm{m}}}=(2 \pi /L,0,0)$ or any of the
two vectors obtained permuting the indexes.

We also study the ferromagnetic properties of the model by monitoring the
usual magnetization
\begin{equation}
  \label{eq:magnet}
  \mathitbf{m} = \frac{1}{V} \sum_{i} \mathitbf{S}_i\,,
\end{equation}
and correspondingly the magnetic susceptibility 
\begin{equation}
  \label{eq:chifm}
  \chi_{m} \equiv V\, \overline{\langle | \mathitbf{m}|^2 \rangle}\,.
\end{equation}
These two observables are crucial to check the possible existence of a
ferromagnetic phase, as predicted by the mean field approximation of
this model~\cite{GROSS}.

\section{Numerical methods}\label{sec:MONTECARLO}

We have analyzed the DPM with $5$ and with $6$ states, on a number of
lattice sizes ($L=4$, $6$, $8$, and $12$).  All the numerical
simulations have been run using a standard Metropolis algorithm
combined with the Parallel Tempering (PT) optimized algorithm, in order
to improve performances and allow to reach thermalization despite the
very large relaxation times typical of spin glass models.

We define a Monte Carlo sweep (MCS) as a set of $V$ trial updates of lattice
spins.  Each simulation consists on a thermalization phase, during which the
system is brought to equilibrium, and a phase of equilibrium dynamics in which
relevant physical observables are measured.  As we require high quality random
numbers, we use a 32-bit Parisi-Rapuano shift register~\cite{PARISI-RAPUANO}
pseudo-random number generator.~\footnote{ Our FPGA did not have components to
  accomodate the L=12 code with a 48 bits generator (that could instead be
  used for L=8).  
We have performed additional numerical simulations in the
  smaller lattices, on PC, using 64 bits random numbers and in the $L=8$, on
  Janus, using 48 bits random numbers.
We have reproduced in all cases, within statistical errors,
  the results obtained with the 32 bits generator.}

In order to improve the simulation performance and to speed up
thermalization we apply a step of the PT algorithm~\cite{PT}
every few MCS's of the Metropolis algorithm.  The PT algorithm is based on the
parallel simulation of various copies of the system, that are governed
by different values of temperature, and on the exchange of their
temperatures according to the algorithm's rules. In practice we let the
different configurations evolve independently for a few MCS, and then we
attempt a temperature swap between all pairs of neighboring
temperatures: the aim is to let each configuration wander in the allowed
temperature range (that goes from low $T$ values, smaller than $T_c$, to
high $T$ values, larger than $T_c$), and to use the decorrelation due to
the high $T$ part of the landscape to achieve a substantial speed up.

In order to check the time scales of the dynamical process, so as to
assess the thermalization and the statistical significance of our
statistical samples, we have computed a number of dynamical observables
that characterize the PT dynamics.

One of them is the temperature-temperature time correlation function,
introduced in Ref.~\onlinecite{Heisenberg}, that we briefly
recall. Let $\beta^{(i)}(t)$ be the inverse temperature of the system $i$ at
time $t$ ($i=0,\ldots,N_T-1$), where $N_T$ is the total number of systems
evolving in parallel in the PT.~\footnote{We have used $\beta$'s not uniformly
  distributed in order to have a PT acceptance of order 30-40\% in the whole
  $\beta$-interval. In addition, we have include additional $\beta$'s in the
  critical region to have  clearer crossing points of the correlation length.}
We consider an arbitrary function of the
system temperature, $f(\beta)$, changing sign at $\beta_\mathrm{c}$. We shall name
$f^{(i)}_t=f(\beta^{(i)}(t))$. In equilibrium, system $i$ can be found at any
of the $N_T$ with uniform probability, hence $\langle
f^{(i)}_t\rangle=\sum_{k=0}^{N_T-1} f(\beta_k) /N_T$, for all $i$ and all
$t$. We must choose a function $f$ as simple as possible, such that $
\sum_{k=0}^{N_T-1} f(\beta_k) =0$.~\footnote{Our choice of $f(\cdot)$ is slightly
  different from that of Ref.~\onlinecite{Heisenberg}; $f(\beta)=a
  (\beta-\beta_\mathrm{c})$ for $\beta<\beta_\mathrm{c}$, and $f(\beta)=b
  (\beta-\beta_\mathrm{c})$ for $\beta > \beta_\mathrm{c}$. The ratio of the
  slopes $a/b$ is fixed by the condition $\sum_{k=0}^{N_T-1}
  f(\beta_k)=0$. The overall normalization being irrelevant, we choose $a=1$.}
Next, we can define the
correlation functions
\begin{equation}
C_f^{(i)}(t)=\frac{1}{N-|t|}\sum_{s=1}^{N-|t|} f_s^{(i)} f_{s+|t|}^{(i)}\,,
\end{equation}
\begin{equation}
\rho_f^{(i)}(t)=\frac{C_f^{(i)}(t)}{C_f^{(i)}(0)}\,,
\end{equation}
where $N$ is the total simulation time.
To gain statistics we consider the sum over all the systems
\begin{equation}
\rho_f(t)=\frac{1}{N_T} \sum_{i=0}^{N_T-1}\rho_f^{(i)}(t)\,.
\label{eq:rho}
\end{equation}
Notice that this correlation function measures correlations for a
given copy of the system, that is characterized, during the dynamics,
by different temperature values. 

We have characterized the correlation function $\rho_f(t)$ through its integrated
autocorrelation time~\cite{VICTORAMIT,sokal96}: 
\begin{equation}
\tau_\mathrm{int}=\int_0^{\Lambda_\mathrm{int}} dt ~\rho_f(t)\, ,
\label{eq:tauint}
\end{equation}
where $\Lambda_\mathrm{int}=\omega\,\tau_\mathrm{int}$ and we have
used $\omega=10$ (we have always used a total simulation time larger
than $15$ or $20$ times $\tau_\mathrm{int}$).

We have studied the systems defined on the smaller lattices ($L=4$ and
$6$) on standard PCs, while for the larger lattice sizes we have used
the Janus computer~\cite{JANUS2,JANUS08}, an FPGA-based machine
specifically designed to handle simulations of spin glass models.  The
performance improvement offered by Janus allowed us to thermalize
lattices of size up to $L=12$.  While the thermalization of lattices
with $L=8$ was relatively fast, the bigger lattice sizes proved to be
rather difficult to equilibrate, even within Janus, things
getting worse as the number of Potts states increases.

Tables~\ref{tab:SIM_DETAILS_p5} and \ref{tab:SIM_DETAILS_p6} summarize
the details about the numerical simulations respectively for the $p=5$
and the $p=6$ case. We were able to thermalize a large number of
samples for $L$ up to $12$. The thermalization of $L=16$ is possible,
but it requires a dramatically large investment in computer resources,
since the time required by each sample is very large. Because of that,
and given the resources we could count upon, we have only been able to
analyze a few samples: the results for the few samples that we have
studied in this case are consistent with the ones obtained from the
smaller sizes.  In addition, for some samples with $L=8$ and $L=12$,
which were especially difficult to thermalize, we had to use larger
numbers of MCS's: see section \ref{sub:THERMALIZATION}.

\begin{table}[ht]
  \begin{center}
    \begin{tabular}{|c|c|c|c|c|c|c|}\hline
      $L$ & $N_\mathrm{samples}$ & MCS$_\mathrm{min}$ &
      $[\beta_\mathrm{min},\beta_\mathrm{max}]$ & $~N_{\beta}~$ & $N_\mathrm{Metropolis}$  &
      $N_m$ \\
      \hline
      \hline
      4  & 2400 & $ 10^7$ & [1.6, 9.5] & 18  & 5 & $10^3$ \\
      \hline                     
      6  & 2400 & $2 \times 10^7$ & [1.6, 9.5] & 22 & 5 & $10^3$ \\
      \hline
      8  & 2448 & $4 \times 10^8$ & [1.7, 6.5] & 24 & 10 & $2 \times 10^5$ \\
      \hline                   
      12 & 2451  & $6 \times 10^9$ & [1.8, 5.5] & 20 & 10  & $2 \times 10^5$ \\
      \hline
    \end{tabular}
  \end{center}
  \caption{Details of the simulations for $p=5$. $N_\mathrm{samples}$
    is the number of samples (i.e. of the disorder realizations that
    we have analyzed), MCS$_\mathrm{min}$ is the minimum number of
    MCSs that we have performed, $[\beta_\mathrm{min},\beta_\mathrm{max}]$ is the
    range of inverse temperatures simulated in the PT,
    $N_\beta$ is the number of temperatures inside this interval,
    $N_\mathrm{Metropolis}$ is the frequency of the Metropolis sweeps per PT step, and
    $N_\mathrm{m}$ is the total number of measurements performed
    within each sample.}\label{tab:SIM_DETAILS_p5}
\end{table}

\begin{table}[ht]
  \begin{center}
    \begin{tabular}{|c|c|c|c|c|c|c|}\hline
      $L$ & $N_\mathrm{samples}$ & MCS$_\mathrm{min}$ & $[\beta_\mathrm{min},\beta_\mathrm{max}]$ & $~N_{\beta}~$ & $N_\mathrm{Metropolis}$  & $N_m$  \\
      \hline
      \hline
      4  & 2400 &  $10^7$ & [2.1, 9.8] & 10 & 5 & $10^3$ \\
      \hline                     
      6  & 2400 &  $2 \times 10^7$ & [2.0, 9.65] & 16 & 5 & $10^3$ \\
      \hline
      8  & 1280 & $10^9$ & [1.7, 7.5] & 30  & 10  & $2 \times 10^5$ \\
      \hline                    
      12 & 1196 &  $6 \times 10^{10}$  & [1.6, 6.5]  & 22 & 10 & $2 \times 10^5$ \\
      \hline
    \end{tabular}
  \end{center}
  \caption{As in table \ref{tab:SIM_DETAILS_p5}, but for $p=6$.}
\label{tab:SIM_DETAILS_p6}
\end{table}

The number of Metropolis sweeps per PT step is $10$ on Janus and $5$ on the
PC, and there is an important reason for that: in a
standard computer the time needed for a step of the PT algorithm is
small compared with a complete Metropolis MCS.  This is not true on Janus,
where it takes longer to perform a PT step than an Metropolis MCS: because of
that, after a careful test of the overall simulation performance, we
decided to lower the PT to Metropolis MCS ratio in order to increase
Janus efficiency.

In the $p=5$ case a numerical simulation of a single sample
(thermalization plus measurements) on Janus takes $39$ minutes for
$L=8$ and $10$ hours on $L=12$.  The same simulations would require
$7.4$ days of an Intel$^{(R)}$ Core2Duo$^{(TM)}$ $2.4$ GHz processor
for $L=8$ and $315$ days for $L=12$. These values grow when $p=6$:
here the equilibration takes $120$ minutes for an $L=8$ sample and
$110$ hours for $L=12$ (on the PC they would take $24$ days for $L=8$
and $10$ years for $L=12$).

The results shown in this paper for the $p=5$ model would have
required approximately $2150$ equivalent years of an Intel$^{(R)}$
Core2Duo$^{(TM)}$ $2.4$ GHz processor: the ones for $p=6$ would have
required $12000$ years.

\section{Results}\label{sec:RESULTS}

\subsection{Thermalization Tests}\label{sub:THERMALIZATION}

Thermalization tests are a crucial component of spin glass
simulations. Before starting to collect relevant results from
the data we have to be sure that they are actually taken
from a properly thermalized system, and are not biased from
spurious effects.

A standard analysis scheme consists in evaluating the average value of
an observable on geometrically increasing time intervals. The whole
set of measurements is divided in subsets, each of which covers only
part of the system's history (the last {\it bin} covers the last half
of the measurements, the previous bin takes the preceding quarter, the
previous bin the previous eighth and so on), and observables are
averaged within each bin. The convergence to equilibrium is checked
comparing the results over different bins: stability in the last three
bins within error bars (that need to be estimated in an accurate way)
is a good indicator of thermalization.

\begin{figure}[ht]
  \includegraphics[width=0.7\columnwidth,angle=270]{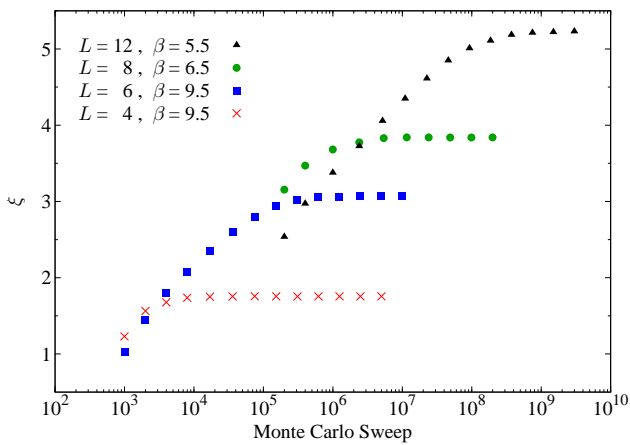}
  \caption{Log-binning thermalization test for $p=5$. For all data
    points the point size is bigger than the corresponding error bar.}
  \label{fig:termp5}
\end{figure}

\begin{figure}[ht]
 \includegraphics[width=0.7\columnwidth,angle=270]{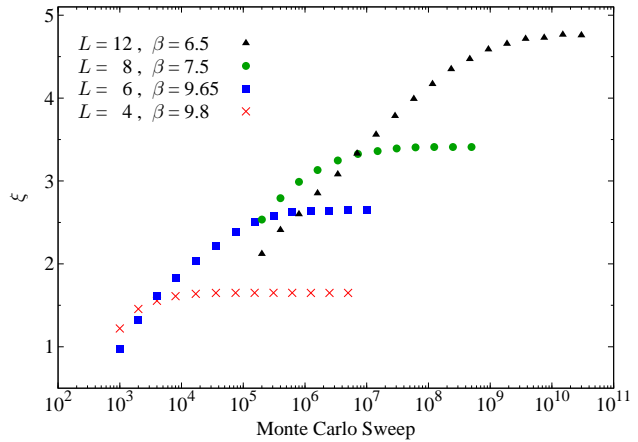}
  \caption{As in figure \ref{fig:termp5}, but $p=6$.}
  \label{fig:termp6}
\end{figure}

We show in figures~\ref{fig:termp5} and~\ref{fig:termp6} the
logarithmic binning of $\xi$, as defined in equation \eqref{eq:xi},
in the $p=5$ and $p=6$ cases. The compatible (and stable) values for
the three last points satisfy the thermalization test explained above.
The data in the plots are for the lowest temperature used on each
lattice size: this is expected to be the slowest mode of the system,
and its thermalization guarantees that also data at higher temperature
values are thermalized.  The plateau in the last part of each plot is
a clear signal of proper thermalization: only data from the last
bin are eventually used to compute thermal averages.

\begin{figure}
  \includegraphics[width=0.7\columnwidth,angle=270,trim=-30 0 40 0]
{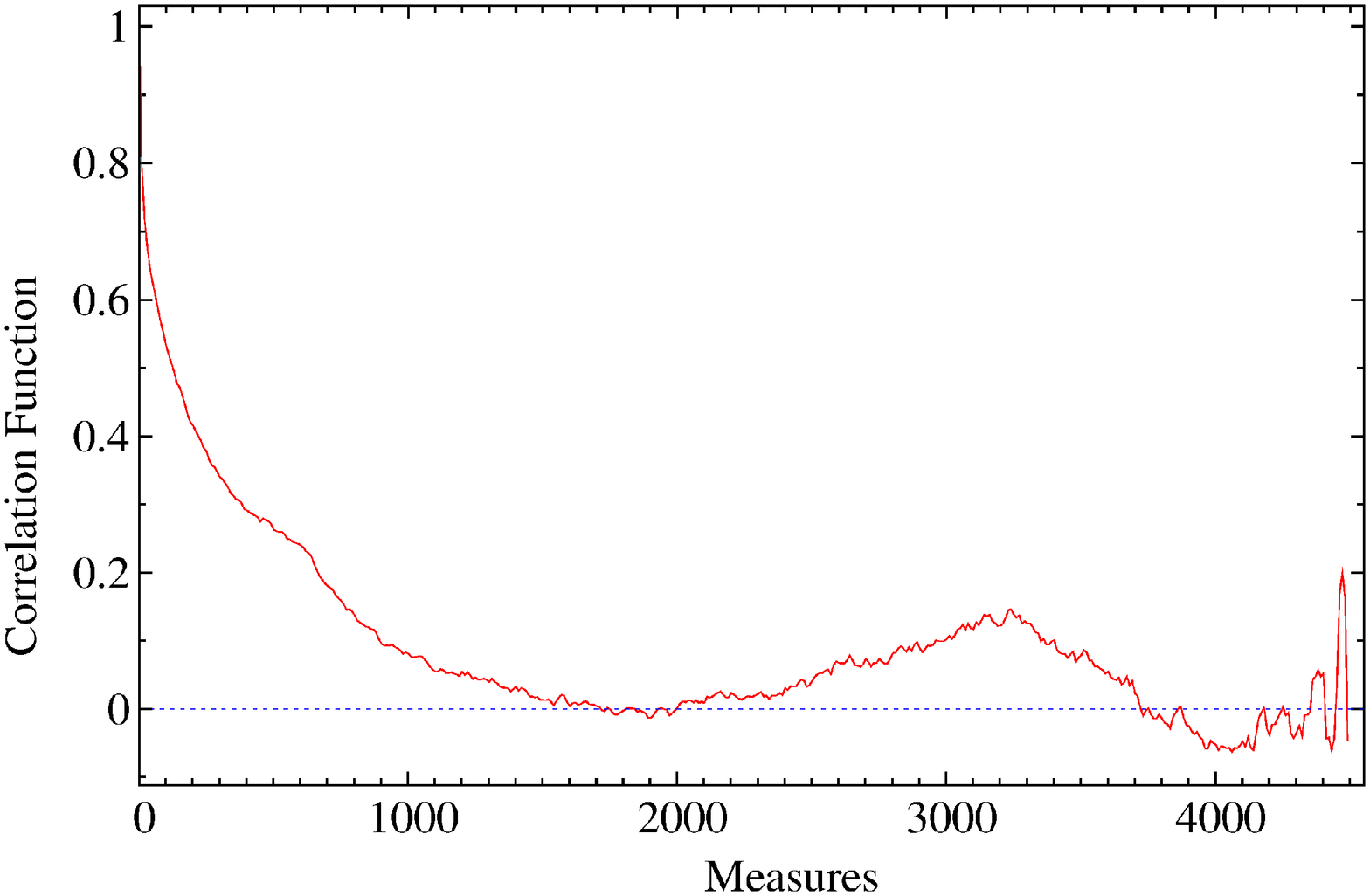}
  \caption{The autocorrelation function \eqref{eq:rho} for one 
   generic sample ($p=6$, $L=8$).}
  \label{fig:correlation_funct}
\end{figure}

We have also investigated how thermalization is reached in the
individual samples (as opposed to the information on averages obtained
from figures \ref{fig:termp5} and \ref{fig:termp6}): to do that we
have studied the correlation function for the temperature random walk
defined in \eqref{eq:rho} and its associated integrated
autocorrelation time, $\tau_\mathrm{int}$, defined in
\eqref{eq:tauint}.  As an example we plot in figure
\ref{fig:correlation_funct} the autocorrelation function
\eqref{eq:rho} for a given sample as a function of the Monte Carlo
time (here $L=8$ and $p=6$): one can see a fast, exponential decay
in the left part of the figure, and (large) fluctuations around zero
at later times.

\begin{figure}
 \includegraphics[width=0.7\columnwidth,angle=270]{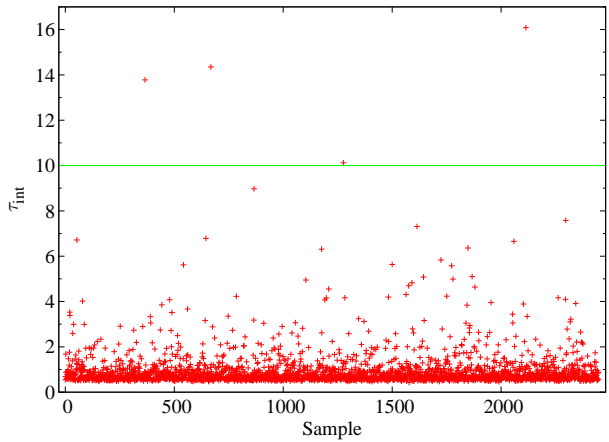}
 \caption{Integrated autocorrelation time, $\tau_\mathrm{int}$, for
   all $p=5$, $L=8$ samples. $\tau_\mathrm{int}$ is in units of blocks
   of ten measurements, i.e. of $20 10^3$ MCS.  Samples above the
   green line have been ``extended'' (see the text for a discussion of
   this issue).}
  \label{fig:tau_int}
\end{figure}

Sample to sample fluctuations of $\tau_\mathrm{int}$ are very large:
in figure~\ref{fig:tau_int} we plot $\tau_\mathrm{int}$ for all our
samples with $p=5$, $L=8$. In order to be on the safe side we have
increased the number of MCS, by continuing the numerical simulation
for a further extent, in all samples where our estimate of
$\tau_\mathrm{int}$ was bigger than the length of the simulation
divided by a constant $c$ ($c=20$ for $L=8$ and $c=15$ for $L=12$,
where achieving thermalization is much more difficult).~\footnote{In
  the $p=5$, $L=8$ case for $2442$ samples we have run a simulation of
  total extent $\eta = 4\times 10^8$ MCSs, while for $5$ samples $\eta
  = 8\times 10^8$ MCSs, and for $1$ sample $\eta = 1.6\times 10^9$
  MCS. In the $p=5$, $L=12$ case for $2382$ samples $\eta = 6\times
  10^9$ MCSs, for $54$ samples $\eta = 1.2 \times 10 ^{10}$, for $8$
  samples $\eta = 2.4 \times 10^{10}$, and for $7$ samples
  $\eta=4.8\times 10^{10}$ MCS. In the $p=6$, $L=8$ case: for $1263$
  samples $\eta=10^9$ MCSs, for $8$ samples $\eta = 2 \times 10^9$ and
  for $9$ samples $\eta = 4\times 10^9$. In the $p=6$, $L=12$ case for
  $1173$ samples $\eta = 6\times 10^{10}$ MCSs, for $17$ samples $\eta
  = 1.2 \times 10^{11}$ MCSs and for $6$ samples $\eta = 2.4 \times
  10^{11}$ MCSs.}

\subsection{Critical temperature and critical exponents}
\label{sub:Tcrit}

Our analysis of the critical exponents of the system has been based on
the quotient method~\cite{VICTORAMIT,QUOTIENT}: by using the averaged
value of a given observable $O$ measured in lattices of different
sizes, we can estimate its leading critical exponent $x_O$,
\begin{equation}
\overline{\langle O(\beta)\rangle} \approx | \beta - \beta_c|^{-x_O} \,.
\label{eq:leadexp}
\end{equation}
By considering two systems on lattices of linear sizes $L$ and $sL$
respectively one has that~\cite{VICTORAMIT,QUOTIENT}
\begin{equation}
\frac{\overline{\langle O(\beta, sL)\rangle}}{\overline{\langle
    O(\beta,L)\rangle}}
=s^{x_O/\nu}+O(L^{-\omega})\,,
\label{QUO}
\end{equation}
where $\nu$ is the critical exponent of the correlation length and
$\omega$ is the exponent of the leading-order
scaling-corrections~\cite{VICTORAMIT}.

We use the operators $\partial_\beta \xi$, from~\eqref{eq:xi}, and
$\chi_q$, from \eqref{eq:chisg} in equation~\eqref{QUO} to obtain
respectively the critical exponents $1+1/\nu$ and $2-\eta_q$.  The
exponent $2-\eta_m$ is obtained applying eq.~\eqref{QUO} to the
magnetic susceptibility $\chi_{m}$, from~\eqref{eq:chifm}.

To use the quotient method we start estimating the
finite-size transition temperature: we do this by looking at the
crossing points of the correlation length in lattice units ($\xi/L$)
for various lattice sizes. We have used a cubic spline interpolating
procedure to compute both the crossings of $\xi/L$ and its
$\beta$-derivative (we have followed the approach described in detail
in Ref.~\onlinecite{POTTSq4}).

\begin{figure}[ht]
  \includegraphics[width=0.7\columnwidth,angle=270]
  {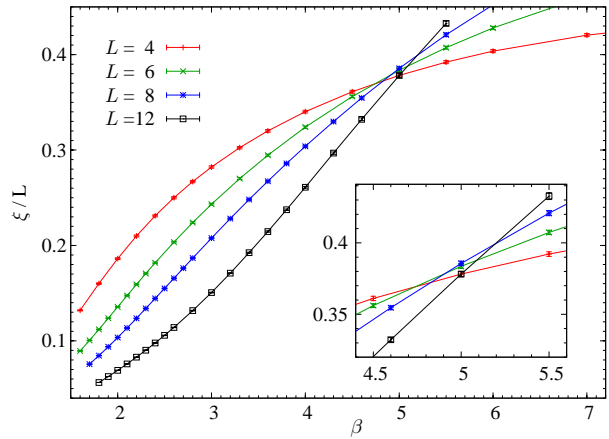}
  \caption{Overlap correlation length in lattice size units as a
    function of the inverse temperature $\beta$ for $L=4$, $6$, $8$
    and $12$. Here $p=5$.}
  \label{fig:cortesp5}
\end{figure}

\begin{figure}[ht]
  \includegraphics[width=0.7\columnwidth,angle=270]
  {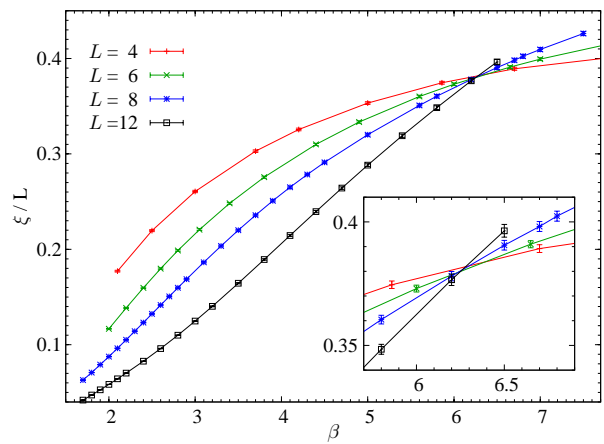}
  \caption{As in figure \ref{fig:cortesp5}, but $p=6$.}
  \label{fig:cortesp6}
\end{figure}

We show in figures~\ref{fig:cortesp5} and~\ref{fig:cortesp6} the
behavior of $\xi/L$ as a function of $\beta$.  The different curves
are for different lattice sizes.  The crossing points are rather clear
in both cases, giving a strong hint of the occurrence of a second
order phase transition. At least for $p=5$ scaling corrections play a
visible role, and the crossing points undergo a small but clear drift
towards lower temperatures for increasing lattice sizes.  We summarize
in tables \ref{tabp5} and \ref{tabp6} the $\beta$ values of the
crossing points for two different pairs of lattice sizes, together
with the estimated values of the critical exponents $\nu$ and $\eta_q$
that we obtain using relation \eqref{QUO}.

Since we can only get reliable results on small and medium size
lattice we cannot control in full scaling corrections, and a
systematic extrapolation to the infinite volume limit is impossible.
It is clear however that the effective critical exponents summarized
in tables~\ref{tabp5} and~\ref{tabp6} do not suggest that
asymptotically for large volume the system will not be critical (in
this case, for example, $\eta_q$ should be asymptotically equal to
$2$): our numerical data clearly support the existence of a finite
temperature phase transition.

\begin{table}[ht]
  \begin {center}
    \begin{tabular}{|c|c|c|c|c|}  \hline 
      $(L_1, L_2)$ & $\beta_\mathrm{cross}(L_1,L_2)$ &$\nu(L_1,L_2)$ & $\eta_q(L_1,L_2)$ &$ \eta_m(L_1,L_2)$ \\ \hline\hline
$(4,8)$ & $4.83(5)$ & $0.82(3)$ & $0.13(2)$ & $1.72(2)$ \\ 
\hline
      $ (6,12)$ & $5.01(4)$ & $0.81(2)$ & $0.16(2)$ & $1.94(2)$  \\
 \hline
    \end{tabular}
  \end{center}
  \caption{Numerical values of our estimates for the crossing point of
    the curves $\xi/L$. We give $\beta_\mathrm{cross}$, the thermal
    critical exponent $\nu$, the anomalous dimension of the overlap
    $\eta_q$, and the anomalous dimension of the magnetization
    $\eta_m$.}
  \label{tabp5}
\end{table}

\begin{table}[ht]
  \begin {center}
    \begin{tabular}{|c|c|c|c|c|}  \hline 
      $(L_1, L_2)$ & $\beta_\mathrm{cross}(L_1,L_2)$ &$\nu(L_1,L_2)$ & $\eta_q(L_1,L_2)$ &$ \eta_m(L_1,L_2)$ \\ \hline\hline
      $(4,8)$ & $6.30(9)$ & $0.80(2)$ & $0.10(2)$ & $1.453(19)$ \\ \hline
$(6,12)$ & $6.26(7)$ & $0.80(4)$ & $0.16(2)$ & $1.971(19)$ \\ \hline
    \end{tabular}
  \end{center}
  \caption{As in table \ref{tabp5}, but $p=6$.}
  \label{tabp6}
\end{table}

We take as our best estimates for the critical exponents the one
obtained from the lattices with sizes $L=6$ and $L=12$. For
$p=5$
\begin{equation}
\beta_c= 5.01(4)\,,\ \nu= 0.81(2)\,,\ \eta_q= 0.16(2)\,,
\end{equation}
while for $p=6$.
\begin{equation}
\beta_c= 6.26(7)\,,\ \nu= 0.80(4)\,,\ \eta_q= 0.16(2)\,.
\end{equation}

It is interesting to compare these values with those of other Potts models
with a different number of states. In particular we are interested in the
value of the critical exponents as a function of the number of states, since
we want to characterize the critical behavior of the various models and
attempt a prediction of the model's behavior when the number of states is
large.  In our particular model and with the (low) values of the temperature that
are interesting for us (since we need to get below the critical point) even
with the large computational power available to us thanks to Janus the
simulation for $p=8$, say, on a $L=12$ lattice, would require an unavailable
amount of CPU time.  What is found in the very interesting work of
Refs. \onlinecite{BBK} and \onlinecite{KLY} is different, since there one is
able to thermalize a $p=10$ model on a large lattice, and no transition is
observed. The model analyzed in these two references~\cite{BBK,KLY} is indeed
slightly (or maybe, it will turn out, not so slightly) different from the
present one, since there $\overline{J}$ is negative. It is not clear to us if
this difference could explain a quite dramatic discrepancy of the observed
behavior, or if, for example, a different (very low) temperature regime should
be analyzed to observe relevant phenomena: this is surely an interesting
question to clarify, and the fact that the coupling have a negative
expectation value, reducing in this way frustration, could turn out to make a
difference.

\subsection{Absence of ferromagnetic ordering in the critical
  region}\label{sub:MAG}

\begin{figure}[ht]
  \includegraphics[width=0.7\columnwidth,angle=270]
{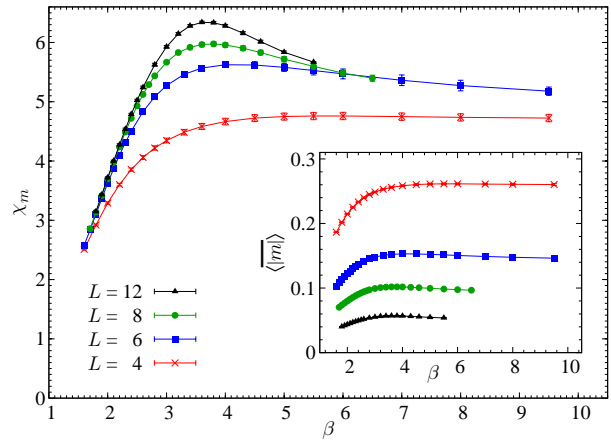}
  \caption{Magnetic susceptibility as a function
    of $\beta$ for $L=4$, $6$, $8$ and $12$. Here $p=5$.}
  \label{fig:sus_magnep5}
\end{figure}

\begin{figure}[ht]
  \includegraphics[width=0.7\columnwidth,angle=270]
{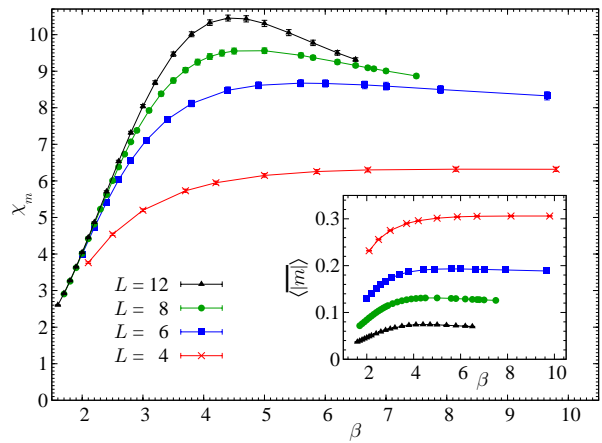}
  \caption{As in figure \ref{fig:sus_magnep5}, but $p=6$.}
  \label{fig:sus_magnep6}
\end{figure}

Our DPM is in principle allowed to undergo a ferromagnetic phase
transition (since no symmetry protects it), and at low temperatures
could present a spontaneous magnetization, as discussed in
Ref.~[\onlinecite{POTTSq4}].  Because of that we have carefully
studied the magnetic behavior of the model at low temperatures.  We
have analyzed both the magnetization and the magnetic susceptibility
below the spin glass critical point.

In the paramagnetic phase the magnetization is random in sign, and its
absolute value is expected to be proportional to $1/\sqrt{V}$. In
Figs.~\ref{fig:sus_magnep5} and~\ref{fig:sus_magnep6} we check whether
$\overline{\langle|\mathitbf{m}|\rangle}$ around the spin
glass
critical
region tends to an asymptotic value for larger lattice size, or not. From the figures
we see $\overline{\langle|\mathitbf{m}|\rangle}$ goes to zero in the critical region.  Also, we
studied the magnetic susceptibility
$\chi_{m}=V\overline{\langle|\mathitbf{m}|^2\rangle}$ which is
independent of size. Again in Figs.~\ref{fig:sus_magnep5}
and~\ref{fig:sus_magnep6} we check that, and we see a non-divergent
behavior.  This behavior is extremely different from a ferromagnetic
phase in which $\chi_{m}$ diverges as the volume.

Besides, as reported in Sec.~\ref{sub:Tcrit} the exponent $\eta_{m}$ is
close to 2, so we could say that a ferromagnetic-paramagnetic phase
transition does not happen in the range of temperatures that we have
studied.

\section{Evolution of critical exponents with $\mathitbf{p}$}\label{crit_exp}

\begin{figure}[ht]
  \includegraphics[width=0.72\columnwidth, angle=270]{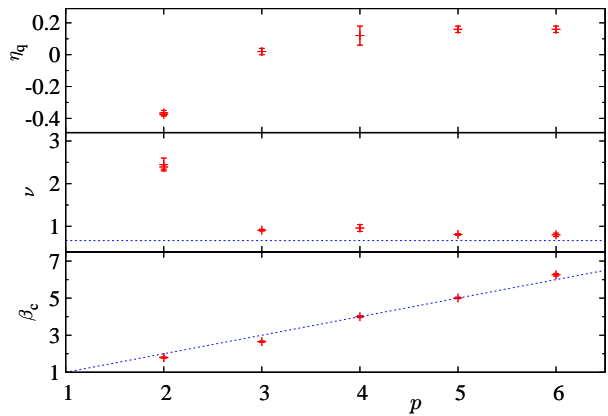}
  \caption{In the bottom plot: $\beta_\mathrm{c}$ versus $p$, and the
    straight line $f(p)=p$. Middle plot: $\nu$ as a function of $p$.
    We also show (dashed line) the value which marks the onset of a
    disordered first order phase transition
    ($\nu_\mathrm{first}=2/3$). Upper plot: $\eta_q$ as a function of
    $p$.}
  \label{fig:nu_p}
\end{figure}

In table~\ref{tab:exp_crit_p} we summarize the values of the inverse
critical temperature and of the thermal and overlap critical exponents
for DPM from $p=2$ (the Ising, Edwards-Anderson spin glass) up to $p=6$.  
We also plot these data items in figure~\ref{fig:nu_p}.

From table~\ref{tab:exp_crit_p} and figure~\ref{fig:nu_p} some results emerge
very clearly. First, the inverse critical temperature roughly follows a linear
behavior in $p$, with a slope very close to one. We have added in
table~\ref{tab:exp_crit_p} the ratio (R) between the numerical determinations
(in $3d$) of $\beta_{\mathrm c}(p)$ and their values in the Mean Field (MF)
approximation. One can see that the large deviations from the MF prediction
occur for large values of $p$ (notice that $R > 1$ since MF suppresses
fluctuations).~\footnote{In the MF approximation was obtained, using the
  Hamiltonian~\cite{GROSS,KW},
$$
  {\cal H} \equiv -\frac{p}{2} \sum_{i \neq j} J_{ij}\, \delta_{s_i,s_j}\,,
$$ that $T_c/J=1$ for $p\le 4$ and $\left( T_c/J \right)^2 =
  1+(p-4)^2/42+O((p-4)^4)$ for $p>4$. In addition for very large $p$, $T_c/J
  \simeq \frac{1}{2}\left(p/\log p \right)^{1/2}$. Taking into account the
  extra $p$ factor in the Hamiltonian used in the Mean Field and the fact that
  $J=\sqrt{2 d}$ ($\overline{J_{ij}^2}=J^2/N$, being $N$ the number of spins
  in the MF computation) since we are working in finite dimension ($d$), we
  obtain the finite dimension version of the critical $\beta$ using the Mean
  Field approximation: $\beta_c=p/\sqrt{2 d}$ for $p\le 4$ and $ \beta_c
  =\frac{p}{\sqrt{2 d}} \left(1-(p-4)^2/84+O((p-4)^4)\right)$ for $p>4$
  (notice the minus signum of the $(p-4)^2$ correction); in addition, for
  large $p$, one obtains $\beta_c\simeq \sqrt{\frac{2}{d}}\left(p \log p
  \right)^{1/2}$. Note that in our case $\sqrt{2 d}\simeq 2.45$.}

Second, $\nu$ decreases monotonically and $\eta_q$ grows monotonically
with the number of states $p$.  To discuss this behavior it is useful to keep
in mind that when using finite size scaling to study a disordered first order
phase transition one expects to find~\cite{DAFF} $\nu=2/D$ and $2-\eta_q=D/2$,
i.e., in our $D=3$ case, $\nu=2/3$ and $\eta_q=1/2$. These are ``effective''
exponents, that are a bound to the ones allowed for second order phase
transitions.

Both sets of values for $\nu$ and $\eta_q$ are indeed completely
compatible with tending, as $p$ increases, to those limit values that
characterize a first order phase transition. If this turns out, as our
numerical data make very plausible to be true, two different
scenarios open.  The first possibility is that the $p$-states DPM
undergoes a disordered first order phase transition for large enough
values of $p$ (just as in the ordered Potts model, that for $p\ge 3$
undergoes a first order phase transition), while the second
possibility is that the DPM will show a standard second order phase
transition for all finite values of $p$. This is the typical issue
that is very difficult to settle with numerical work: an analytical
solution of the model with infinite number of states would be very
useful as a starting point in order to discriminate between these two
possible scenarios.

\begin{table}[ht]
  \begin{center}
    \begin{tabular}{|c|c|c|c|c|}\hline
      $ p $ & $\beta_\mathrm{c}$ & $\nu$ & $\eta_{q}$ & $R$ \\
      \hline
      \hline
$ 2 $ (Ref.[\onlinecite{KKY}])  &      $1.786(6)$ &
$2.39(5)$ \footnote{This value of $\nu$ is from $\xi_L/L$. It is different and
  more reliable than the one obtained from the spin glass
  susceptibility, that, because of large scaling corrections, would
  severely depend on the kind of analysis.}   & 
$-0.366(16)$\footnote{This value of $\eta_q$ is from the study of the spin glass
  susceptibility.}& 2.187(8) \\  \hline
$ 2 $ (Ref.[\onlinecite{PELI}]) &      $1.804(16)$  & $2.45(15)$  &
      $-0.375(10)$& 2.209(20)  \\\hline
$ 3 $ (Ref.[\onlinecite{KLY}]) &       $2.653(35)$  & $0.91(2)$   & $0.02(2) $
      & 2.17(3)\\\hline
$ 4 $ (Ref.[\onlinecite{POTTSq4}]) &   $4.000(48)$  & $0.96(8)$   & $0.12(6)$&
      2.45(3)\\\hline
$5$ (this paper) &                          $5.010(40) $  & $0.81(2) $  &
      $0.16(2) $& 2.51(2) \\ \hline
$6$ (this paper) &                          $6.262(71) $  & $0.80(4)$   &
      $0.16(2)$ & 2.69(3) \\\hline
     \end{tabular}
  \end{center}
  \caption{Critical parameters as a function of $p$. All data are for
    binary couplings, with zero expectation value. By $R$ we denote the ratio
    between the critical $\beta$ in three dimensions and that computed in Mean
  Field.
}
\label{tab:exp_crit_p}
\end{table}

\section{Conclusions}\label{CONCLUSIONS}

In this note we have characterized the critical behavior of the $3D$
DPM with $p=5$ and $p=6$, i.e. with a reasonably large number of
states.  Our numerical simulations have allowed us to reach some clear
evidences, and to stress some difficult issues that will require
further analysis.

We first stress that in both cases the spin glass transition is very clear,
and we have been able to obtain a reliable estimate of the critical
temperature and of the critical exponents $\nu$ and $\eta_q$.  We have
discussed what happens when $p$ increases; we have found that $\beta_c$
increases like $p$. A similar result was conjectured in Ref.  \onlinecite{HTE}
(for all values of $p$) analyzing high temperature series and found in Mean
Field for $p\le 4$ (although, of course, the slope is wrong).  
In addition, the behavior
of $\nu$ and $\eta_q$ is compatible with going to the large $p$ limit value
that characterizes a first order phase transition.

In the low temperature regime we do not see any sign of a transition
to a ferromagnetic regime, that would be in principle allowed by the
structure of our model. We cannot exclude that at very low $T$ values
something would happen, but in all the range we can explore the system
stays in the spin glass phase. 

A last piece of important evidence is that low temperature simulations
of this model look difficult, and that they slow down severely for
increasing $p$. In our particular model, where the expectation of the
coupling is zero, it would be impossible to study reliably a $p=8$
model with the computational resources available today.

This last observations opens indeed a last point that it will be
interesting to analyze in the future. When couplings have a negative
expectation value the simulation of a $p=10$ model ~\cite{BBK,KLY} is
possibly easier than it would be in our case, and the results are very
different: in that case one does not see any sign of a phase transition.
Analyzing how the DPM depends on the expectation value of the couplings
is indeed at this point a crucial issue, since it could turn out that
the reduction in frustration due to a negative net value of the
couplings could completely change the critical behavior of the model.

\section*{Acknowledgments}

Janus has been funded by European Union (FEDER) funds, Diputaci\'on General de
Arag\'on (Spain), by a Microsoft Award - Sapienza - Italy, and by Eurotech. We
were partially supported by MICINN (Spain), through contracts
No. TEC2007-64188, FIS2006-08533-C03, FIS2007-60977, FIS2009-12648-C03 and
UCM-Banco de Santander. D. Yllanes and B. Seoane are FPU Fellow (Spain).  S.P.-G. was
supported by FECYT (Spain). The authors would like to thank the Ar\'enaire
team, especially J. Detrey and F. de Dinechin for the VHDL code of the
logarithm function\cite{log}.


\end{document}